\documentclass[aps,prc,reprint,showpacs,floatfix,superscriptaddress]{revtex4-2}
\usepackage[colorlinks=true,linkcolor=blue]{hyperref}%
\usepackage{physics}
\usepackage{graphicx}
\usepackage{lipsum}
\usepackage{mwe}
\usepackage{caption}
\usepackage{dcolumn}
\usepackage{bm}
\usepackage{adjustbox}
\usepackage{rotating}
\usepackage{xcolor}
\usepackage{ulem}
\usepackage{tikz}

\begin{document}
\preprint{APS/123-QED}
\title{Hot pygmy dipole strength in nickel isotopes} 

\author{Amandeep Kaur}\email[]{akaur.phy@pmf.hr}
\affiliation{Department of Physics, Faculty of Science, University of Zagreb, Bijeni\v{c}ka c. 32,  10000 Zagreb, Croatia}

\author{Esra Y{\"{u}}ksel}\email[]{e.yuksel@surrey.ac.uk}
\affiliation{School of Mathematics and Physics, University of Surrey, Guildford, Surrey GU2 7XH, United Kingdom}

\author{Nils Paar}\email[]{npaar@phy.hr}
\affiliation{Department of Physics, Faculty of Science, University of Zagreb, Bijeni\v{c}ka c. 32,  10000 Zagreb, Croatia}

\date{\today}

\begin{abstract} 
At finite temperatures, nuclear excitations are significantly modified, most notably through the emergence of additional low-energy dipole strength, which can critically impact astrophysical reaction rates. Ongoing fusion-evaporation experiments on Ni isotopes provide a unique opportunity to investigate the hot pygmy dipole strength (HPDS), underscoring the need for reliable theoretical predictions and a comprehensive understanding of this emerging phenomenon. In this work, the HPDS is investigated in Ni isotopes from $N = Z$ to neutron-rich systems ($^{56\text{--}70}$Ni) over a temperature range of $T=$ 0$-$2~MeV using the finite-temperature relativistic quasiparticle random phase approximation. In neutron-rich Ni isotopes, the pygmy dipole strength at higher temperatures exceeds up to 2.5 times its value observed at zero temperature. In contrast, near $N \approx Z$ isotopes show negligible low-energy dipole strength at $T = 0$ MeV but develop a pronounced HPDS as the temperature increases. Predicted E1 energy-weighted strength ($S_{\text{EWS}}$) and cumulative $B$(E1) values for HPDS are presented across the Ni isotopic chain for various low-energy intervals and temperatures, providing essential benchmarks to support and guide experimental studies.
\end{abstract}

\maketitle

\textit {Introduction.} \label{Intro}
The pygmy dipole strength (PDS), often referred to as the pygmy dipole resonance (PDR), has attracted considerable interest, both experimentally \cite{SAVRAN2013210, BRACCO2019360} and theoretically \cite{Paar_2007,LANZA2023104006,PhysRevC.89.041601,paarproperties}, due to its sensitivity to the isospin properties of nuclei, including the neutron skin thickness, as well as its potential impact on astrophysical processes in stellar environments~\cite{PhysRevC.76.051603,LITVINOVA200926,GORIELY2002217,PhysRevC.81.041301,GORIELY2004331}. In particular, the enhanced $\gamma$-ray strength function associated with the PDS can significantly influence radiative neutron-capture $(n,\gamma)$ reaction rates, which play a crucial role in the rapid neutron-capture process ($r$-process) \cite{GORIELY2004331,PhysRevC.82.014318,Paar_2007,PhysRevC.109.054311}. This process is responsible for the synthesis of heavy elements in extreme astrophysical sites, such as core-collapse supernovae and neutron star mergers \cite{siegel2022r,ARNOULD200797, RevModPhys.93.015002}.

Typically found below the giant dipole resonance (GDR) region, the PDS is characterized by an enhancement of the electric dipole (E1) strength near the particle separation energy, resulting from a rather complex interplay of excitations involving excess neutrons, as well as some proton contributions ~\cite{RocaMaza2012,Vretenar2012,BRACCO2019360}. While the GDR dominates the E1 response and nearly exhausts the Thomas--Reiche--Kuhn (TRK) sum rule, the PDS is known to contribute only a modest fraction of the total dipole strength \cite{Paar_2007,SAVRAN2013210,BRACCO2019360}. Due to its strong correlation with both the neutron skin thickness and the electric dipole polarizability \cite{PhysRevC.83.034319,PhysRevC.88.044610,Paar2014,PhysRevC.92.064304,YUKSEL2023137622}, the PDS also serves as a sensitive probe of the isovector sector of the nuclear equation of state (EoS), which is directly linked to the density dependence of the nuclear symmetry energy---a key quantity governing the properties of neutron-rich matter, including neutron stars and their mergers \cite{Lattimer2012,Oertel2017,PhysRevC.90.011304}. The nearly linear correlation between the PDS, dipole polarizability, and neutron skin thickness makes it a valuable observable for constraining symmetry energy parameters~\cite{PhysRevC.76.051603,PhysRevC.85.051601,PhysRevC.84.027301,PhysRevC.81.041301,ROCAMAZA201896,bertulani2019pygmy,YUKSEL2023137622}. 

In this context, neutron-rich nickel (Ni) isotopes provide benchmark systems for experimental investigations into the evolution of the PDS as a function of increasing neutron-to-proton asymmetry. For instance, the virtual photon scattering method has been employed to investigate the PDS in $^{68}$Ni, yielding a measured low-lying E1 energy-weighted sum rule strength ($S_{\text{EWS}}$) of approximately 5\% of the classical TRK sum rule ~\cite{PhysRevLett.102.092502}. In a separate experiment using the Coulomb excitation method, the PDS in $^{68}$Ni was found to exhaust 2.8\% of the TRK sum rule at an excitation energy of $E = 9.55$~MeV~\cite{PhysRevLett.111.242503}. For $^{70}$Ni, the low-lying E1 strength was measured as $S_{\text{EWSR}} = (1.5\% \pm 0.2\%)$ TRK in the energy range $E = 6\text{--}8$~MeV and $(4.8\% \pm 0.9\%)$ TRK in the range $E = 8\text{--}12$~MeV~\cite{PhysRevC.98.064313}. 

In addition to isospin asymmetry, finite temperature can significantly influence both the overall properties and the underlying structure of the PDS in nuclei~\cite{NIU2009315}. While the properties of the GDR under extreme temperature ($T$) conditions  have been extensively studied and measured~\cite{santonocito2020hot,doi:10.1142/S0217732307024474,PhysRevLett.97.012501}, the behavior of the PDS at finite temperature remains largely unexplored. To date, no experimental data on the \textit{hot} pygmy dipole strength (HPDS) have been available. However, recent experiments employing fusion-evaporation reactions aim to investigate the HPDS in Ni isotopes, and preliminary results provide the first evidence of its existence~\cite{wieland2024extra,wieland2025extra,Wieland2023,Wieland2024}. These ongoing studies are expected to provide valuable insights and quantitative data on the $T$-dependence of pygmy dipole excitations in the near future, shedding light on their role in nuclear structure and their broader implications in astrophysical environments.

Several theoretical investigations of the PDS at zero temperature, based on both relativistic and non-relativistic formulations of the random phase approximation (RPA), have highlighted the role of neutron excess in the emergence of low-energy dipole modes in nuclei with extreme neutron-to-proton ($N/Z$) ratios \cite{Paar_2007, LANZA2023104006,PhysRevC.85.044317,VRETENAR2001496,PAAR2005288}. At finite temperature, the behavior of low-energy monopole and dipole excitations has been studied using finite-temperature RPA (FT-RPA), revealing that thermal unblocking of single-particle orbitals near the Fermi surface enhances the low-energy transition strength \cite{NIU2009315}. A similar temperature-induced enhancement of isovector low-energy dipole strength has been confirmed in finite-temperature quasiparticle RPA (FT-QRPA) calculations \cite{Yuksel55,PhysRevC.96.024303}, as well as in finite-temperature continuum QRPA frameworks based on Skyrme energy density functionals \cite{PhysRevC.88.031302,KHAN2004311}. Furthermore, the finite-temperature relativistic time-blocking approximation (FT-RTBA), developed within the Matsubara Green’s function formalism using the time-blocking method, enables investigation of both low- and high-energy E1 strength distributions in thermally excited nuclei \cite{PhysRevC.100.024307}.

Motivated by ongoing experimental investigations of the HPDS, this work explores its temperature-dependent evolution in nickel isotopes ($^{56\text{--}70}$Ni), spanning from symmetric to neutron-rich systems. The temperature interval $T = 0$ to 2 MeV is chosen to align with the range explored in ongoing experimental investigations. Calculations are performed within the framework of the finite-temperature relativistic quasiparticle random phase approximation (FT-RQRPA)~\cite{PhysRevC.109.014314,PhysRevC.109.024305}, based on a relativistic energy density functional (REDF). This approach enables a fully self-consistent description of temperature effects on nuclear excitations, accounting for modifications in both single-particle spectra and residual interactions. Our study focuses on the emergence and evolution of the HPDS, offering insights into the thermal response of exotic nuclei and the interplay between temperature and neutron excess, as well as their impact on the underlying structure of the HPDS. To enable direct comparison with forthcoming experimental data, we present a detailed table summarizing the calculated properties of the HPDS.

This letter first summarizes the theoretical framework of the recently developed FT-RQRPA, and followed by the results for the HPDS in the Ni isotopic chain, with a detailed discussion of the impact of neutron excess and temperature on low-energy isovector dipole excitations. The letter concludes with a summary of the main findings, highlighting the emergence of HPDS as a temperature-dependent phenomenon, and discusses their relevance to ongoing and future experimental investigations. 

\textit{Theoretical framework.} 
To investigate the low-energy E1 strength in Ni isotopes at finite temperature, we employ the recently developed finite-temperature relativistic quasiparticle random phase approximation (FT-RQRPA) \cite{PhysRevC.109.014314,PhysRevC.109.024305}, which is based on the relativistic energy density functional (REDF) with the point-coupling interaction DD-PCX \cite{PhysRevC.99.034318}. A brief overview of the theoretical framework is provided here and further details can be found in Refs.~\cite{PhysRevC.109.014314,PhysRevC.109.024305,GOODMAN198130}.
We note that the implementation of DD-PCX effective interaction, which explicitly includes density dependence in the vertex functionals, is essential for considerations of E1 transition strength in nuclei. Density dependent functionals provide reasonable description of the symmetry energy at saturation density $(J)$ \cite{ROCAMAZA201896} (for DD-PCX, $J=$~31.12 MeV \cite{PhysRevC.99.034318}), which governs the dipole transition strength. In contrast, non-linear meson exchange interactions, having large symmetry energy at saturation density, e.g., for NL3$^{*}$, $J=$ 38.68 MeV \cite{LALAZISSIS200936} or for NL3, $J=$~37.4 MeV \cite{PhysRevC.55.540}, result in overestimated E1 transition strengths, including those in low-energy region \cite{PhysRevC.98.064313,PhysRevC.109.054311}.
\begin{figure*}[ht]
    \resizebox{!}{0.54\textwidth}{
    \includegraphics{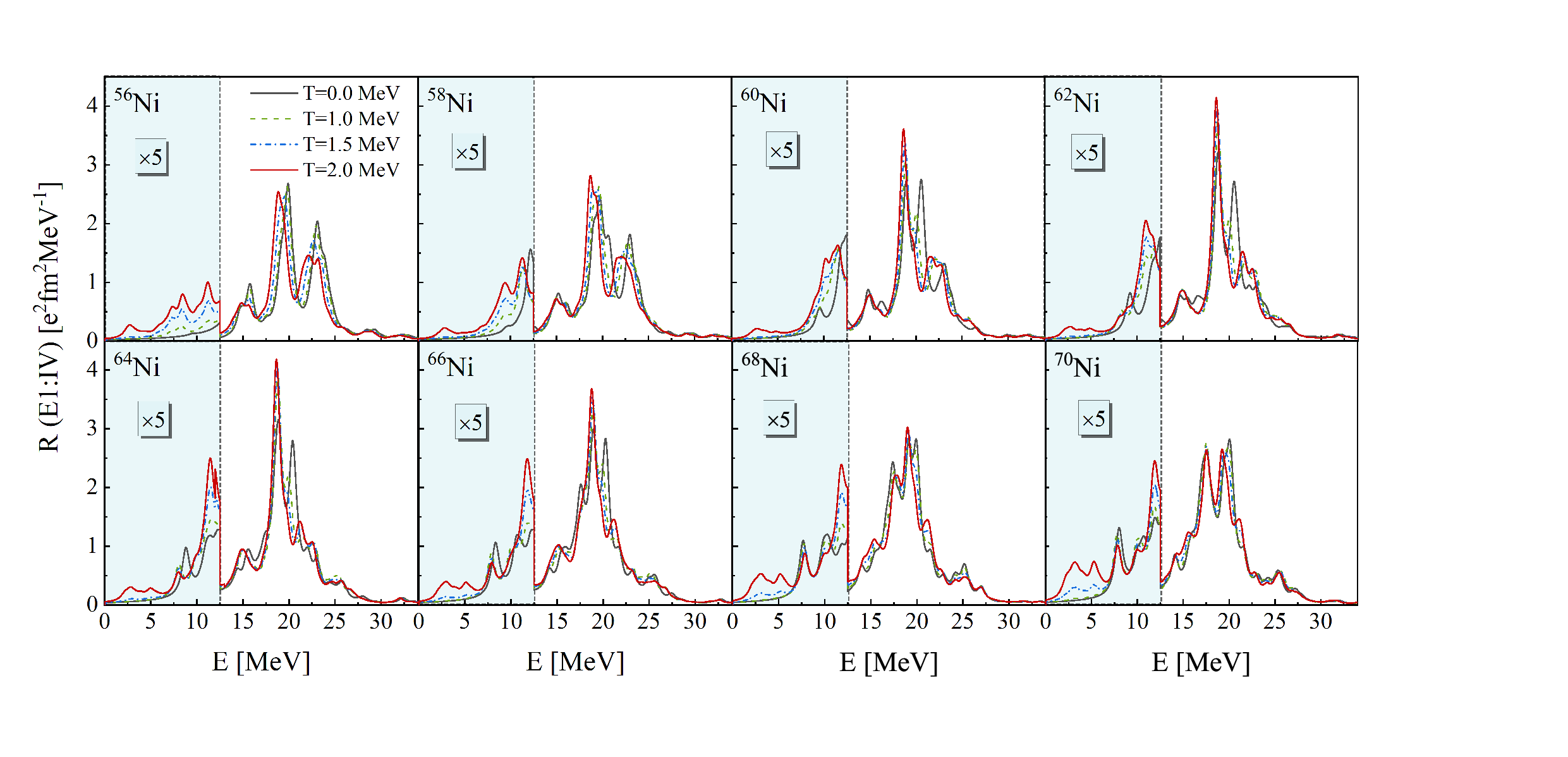}} \vspace{-1.5cm}
    \caption{The isovector E1 strength distributions for $^{56-70}$Ni calculated at $T$ = 0, 1, 1.5 and 2 MeV using FT-RQRPA. The light blue color represents the PDS strength below 12 MeV, magnified by a factor of five for visibility.}
    \label{fig1}
\end{figure*}

The nuclear properties are described using the finite-temperature Hartree–Bardeen–Cooper–Schrieffer (FT-HBCS) approach \cite{GOODMAN198130,yuksel2014effect}. The $T$-dependent Fermi-Dirac distribution function is given by
\begin{equation}
\label{fermi_dirac}
f_{i}=[1+exp(E_{i}/k_{B}T)]^{-1},
\end{equation}
where $k_{B}$ and $T$ denote the Boltzmann constant and temperature, respectively. The quasiparticle (q.p.) energy $E_i$ of a state is calculated using the expression $E_i = \sqrt{(\varepsilon_i - \lambda_q)^2 + \Delta_i^2}$, where $\varepsilon_i$ represents the single-particle energy, $\lambda_q$ is the chemical potential for either protons or neutrons, and $\Delta_i$ denotes the pairing gap of the corresponding state. A sharp pairing phase transition from the superfluid to the normal state is expected at critical temperature ($T_c$) \cite{GOODMAN198130}. The values of $T_c$ for the open-shell Ni isotopes are calculated using the FT-HBCS approach and are given in Table \ref{table1}.

\begin{table}[htp] \renewcommand{\arraystretch}{1.2}
\tabcolsep 0.25cm
\caption{ The critical temperature ($T_c$) values for pairing phase transition in open-shell Ni isotopes.}
\centering 
\begin{tabular}{c c c c } 
\hline \\[-1.0em]
 Nucleus   & $T_c$ [MeV] & Nucleus  & $T_c$ [MeV] \\
\hline                                         
$^{58}$Ni  &  0.69     & $^{64}$Ni      & 0.89   \\
$^{60}$Ni  &  0.83     & $^{66}$Ni      & 0.78   \\
$^{62}$Ni  &  0.89     & $^{70}$Ni      & 0.64   \\
\hline \\ [-1.ex]
\end{tabular}
\label{table1} 
\end{table}

Using the Fermi-Dirac distribution function given in Eq.~(\ref{fermi_dirac}), the $T$-dependent occupation probabilities of single-particle states are expressed as
\begin{equation}
n_i=v_{i}^{2}(1-f_{i})+u_{i}^{2}f_{i},
\end{equation} 
where $u_i$ and $v_i$ are the BCS amplitudes. The excitation operator and the derivation of the expressions for its matrix elements are detailed in Refs. \cite{Yuksel55,PhysRevC.96.024303}. The FT-RQRPA matrix is given as 
\begin{equation}
\left( { \begin{array}{cccc}\label{eq:qrpa}
 \widetilde{C} & \widetilde{a} & \widetilde{b} & \widetilde{D} \\
 \widetilde{a}^{+} & \widetilde{A} & \widetilde{B} & \widetilde{b}^{T} \\
-\widetilde{b}^{+} & -\widetilde{B}^{\ast} & -\widetilde{A}^{\ast}& -\widetilde{a}^{T}\\
-\widetilde{D}^{\ast} & -\widetilde{b}^{\ast} & -\widetilde{a}^{\ast} & -\widetilde{C}^{\ast}
 \end{array} } \right)
 \left( {\begin{array}{cc}
\widetilde{P}  \\
\widetilde{X }  \\
\widetilde{Y}  \\
\widetilde{Q} 
 \end{array} } \right)
 = E_{w}
  \left( {\begin{array}{cc}
\widetilde{P}  \\
\widetilde{X}  \\
\widetilde{Y}  \\
\widetilde{Q} 
\end{array} } \right), \end{equation}
where $ E_{w}$ denotes the excitation energies, and $\widetilde{P}$, $\widetilde{X}$, $\widetilde{Y}$, $\widetilde{Q}$ are eigenvectors.
The FT-RQRPA matrices are diagonalized in a self-consistent way, allowing a state-by-state analysis for each excitation. The detailed description of the $T$-dependent matrix elements is given in \cite{PhysRevC.96.024303,SOMMERMANN1983163}. At finite temperature, the reduced transition probability is calculated as
%
\begin{equation}
\begin{split}
B(EJ,\widetilde0\rightarrow w)&=\bigl|\langle w ||\hat{F}_{J}||\widetilde0\rangle \bigr|^{2}\\
&=\biggl|\sum_{c\geq d}\Big\{(\widetilde{X}_{cd}^{w} + (-1)^{j_{c}-j_{d}+J}\widetilde{Y}_{cd}^{w}) \\
&\times(u_{c}v_{d}+(-1)^{J}v_{c}u_{d})\sqrt{1-f_{c}-f_{d}} \\
&+(\widetilde{P}_{cd}^{w}+(-1)^{j_{c}-j_{d}+J}\widetilde{Q}_{cd}^{w})\\
&\times(u_{c}u_{d}-(-1)^{J}v_{c}v_{d})\sqrt{f_{d}-f_{c}}\Big\}\langle c ||\hat{F}_{J}||d\rangle\biggr|^{2}.
\end{split}
\label{bel}
\end{equation}
%
$|w\rangle$ denotes the excited state and $|\widetilde0\rangle$ is the correlated FT-RQRPA vacuum state. $\hat{F}_{J}$ is the transition operator of the relevant excitation. In this work, the isovector E1 operator is used to calculate electric transition strength distributions \cite{PhysRevC.67.034312}. It would be insightful to examine the contribution of a specific neutron or proton configuration to the total E1 transition strength at a given excitation energy $E_w$, 
\begin{equation}\label{Part.Contri}
B (EJ, E_{w})=\big|\sum_{cd}\left(b^{\pi}_{cd}(E_w)+b^{\nu}_{cd}(E_w)\right)\big|^2.
\end{equation}
Here, \( b^{\pi}_{cd}(E_w) \) and \( b^{\nu}_{cd}(E_w) \) represent the proton (\( \pi \)) and neutron (\( \nu \)) partial contributions for a specific configuration \( cd \). Finally, the discrete FT-RQRPA spectrum is smoothed using a Lorentzian averaging with a width of \( \Gamma = 1.0 \) MeV, according to the following expression,
\begin{equation}\label{lorentz}
R(EJ,E_{w})=\sum_{w}\frac{1}{2\pi}\frac{\Gamma}{(E-E_{w})^{2}-\Gamma^{2}/4}B(EJ,\widetilde0\rightarrow w).
\end{equation}

\textit{Results and discussion.} 
First, we present the FT-RQRPA analysis of the evolution of the isovector (IV) dipole response in both low- and high-energy regions along the Ni isotopic chain at zero and finite temperatures. Fig.~\ref{fig1} shows the isovector E1 strength distributions for the $^{56\text{--}70}$Ni isotopes at $T = 0$, 1, 1.5, and 2~MeV, calculated using the FT-RQRPA with DD-PCX interaction. The low-energy region is magnified by a factor of five and highlighted in light blue to enhance visibility. At \( T = 0 \)~MeV, the doubly magic nucleus $^{56}$Ni exhibits no evidence of PDS. However, with increasing neutron number along the Ni isotopic chain, a pronounced enhancement of the low-energy dipole strength is observed below $E = 12$~MeV. These low-lying dipole states predominantly arise from transitions involving valence neutrons. 

To benchmark FT-RQRPA calculations in the \( T = 0 \) limit, which corresponds to the RQRPA, Table~\ref{table2} presents the \( S_{\text{EWS}} \) values, expressed as a percentage of the TRK sum rule for the \( E = 0\text{--}12 \)~MeV region. These results are compared with the experimental data available for $^{68}$Ni and $^{70}$Ni ~\cite{PhysRevC.98.064313,PhysRevLett.111.242503,PhysRevLett.102.092502}. To facilitate comparison with our results, the experimental $S_{\text{EWS}}$ values and their associated uncertainties were integrated over the energy range $E = 0$–12~MeV. The RQRPA results for $^{68}$Ni show very good agreement with the experimental data; however, the calculations slightly underestimate the values for $^{70}$Ni. Nonetheless, the overall results remain reasonably consistent with the experimental values for both isotopes, thereby supporting the extension of the calculations to finite temperatures.

\begin{table}[htp!] \renewcommand{\arraystretch}{1.2}
\tabcolsep 0.15cm
\caption{The calculated
$S_{\text{EWS}}$ with respect to the TRK sum rule for the energy region $E = 0$--12~MeV in $^{68,70}$Ni, compared with available experimental data.}
\centering 
\begin{tabular}{c c c} 
\hline \\[-1.0em]
              & RQRPA   & Exp.             \\
\hline                                         
$^{68}$Ni     & 3.1\%       & 2.8\%$\pm$0.5\% \cite{PhysRevLett.111.242503}; 5.0\%$\pm$1.5\% \cite{PhysRevLett.102.092502}        \\ 
$^{70}$Ni     & 3.7\%       & 6.3\%$\pm$1.1\% \cite{PhysRevC.98.064313}  \\
\hline \\ [-1.ex]
\end{tabular}
\label{table2} 
\end{table}

\begin{figure}[htp!]
\includegraphics[width=\linewidth,clip=true]{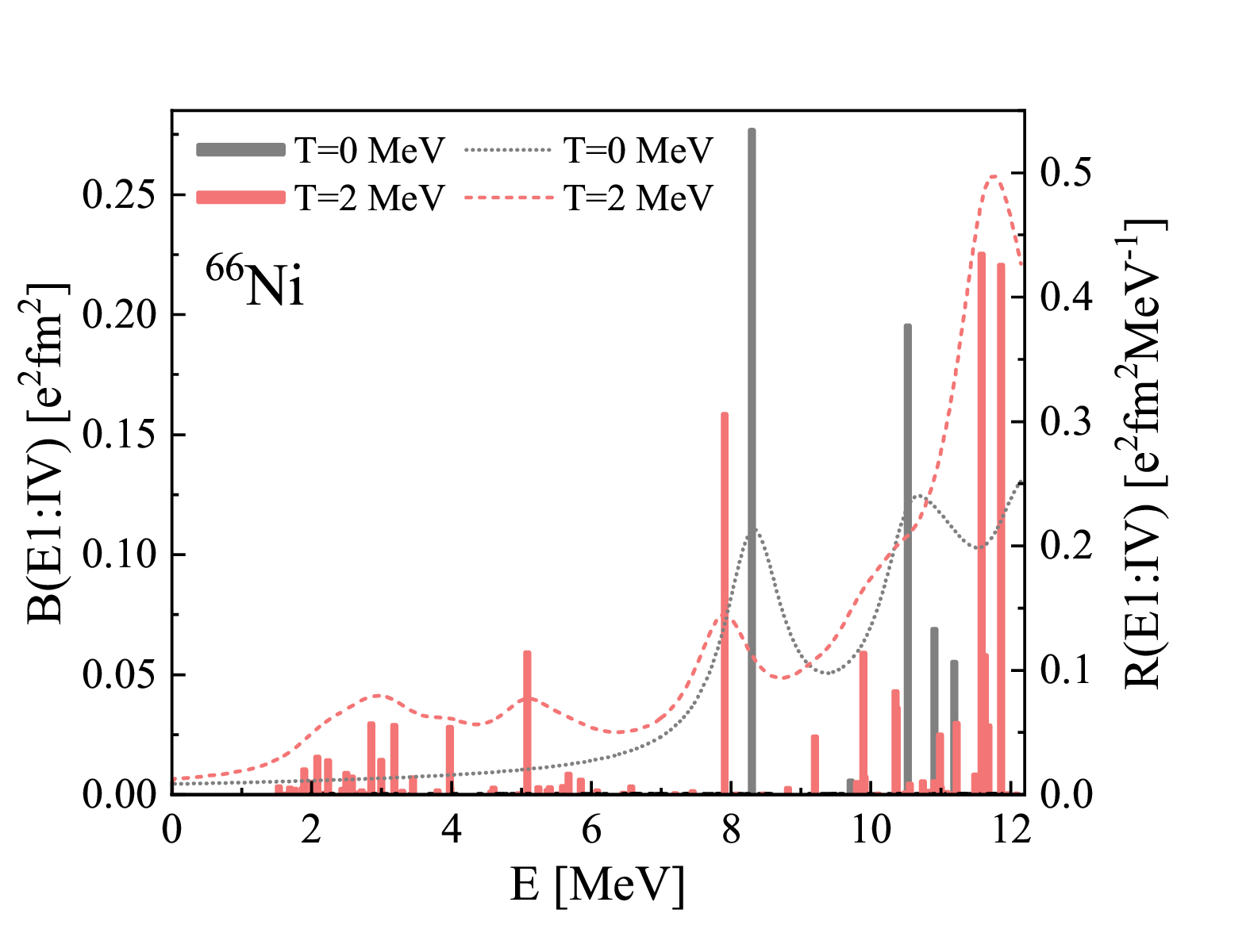}
  \vspace{-0.7cm}
  \caption{Isovector E1 transition probability B(E1) (vertical bars)
  and strength function R(E1) (dashed/dotted lines) for $^{66}$Ni in the low-energy region at $T=$ 0 and 2 MeV.}
  \label{fig2} 
\end{figure}

\begin{figure*}[ht]
    \resizebox{!}{0.4\textwidth}{
    \includegraphics{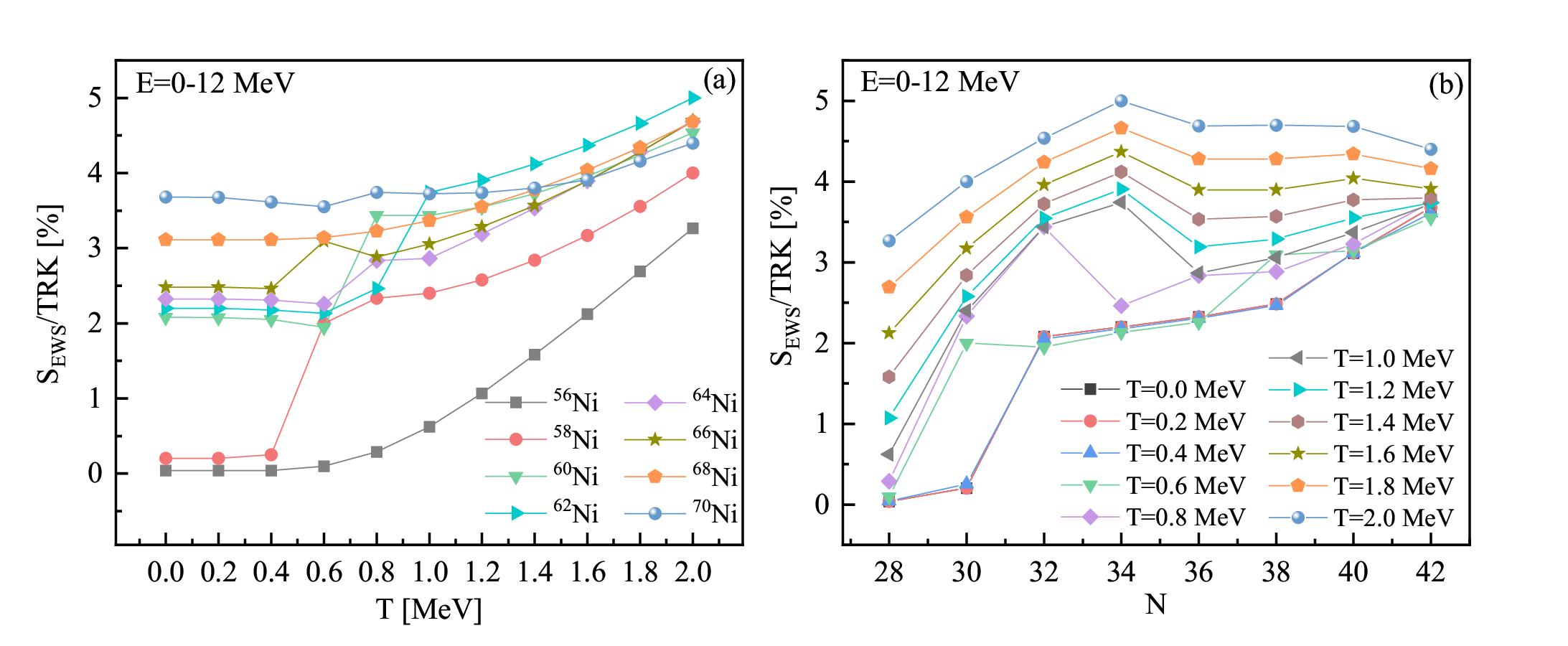}} \vspace{-0.4cm}
    \caption{For the low-energy region $E=0-12$ MeV, the evolution of the $S_{EWS}$ (\%) with respect to TRK sum rule in Ni isotopes is shown as a function of (a) neutron number (N) and (b) temperature (T).}
    \label{fig3}
\end{figure*}

In the temperature range from $T = 1$ to 2~MeV, the high-energy E1 strength is redistributed across the main peaks, and the associated excited states gradually shift towards lower energies, as shown in Fig.~\ref{fig1}. The effects of temperature are even more pronounced in the low-energy region. Ni isotopes with lower neutron content also begin to exhibit the emergence of PDS, with the most drastic changes observed in $^{56}$Ni. At $T = 1.5$~MeV, new low-energy states begin to appear below $E \approx 6$~MeV. As the temperature increases further to $T = 2$~MeV, this effect becomes even more prominent, particularly in neutron-rich nuclei, where additional low-energy excited states with appreciable strength emerge.
For clarity, the reduced transition probability $B$(E1) (shown as vertical bars) and the dipole strength function $R$(E1) (represented by dashed and dotted lines) for $^{66}$Ni are plotted in Fig.~\ref{fig2}, focusing on the low-energy region.
The figure clearly illustrates that the impact of temperature on E1 transitions is most pronounced in the low-energy region. Notably, at $T = 2$~MeV, several new excited states emerge below 8~MeV, which are absent at zero temperature. As the temperature increases, nucleons become thermally excited and begin to populate higher-lying single-particle states. This thermal promotion leads to changes in occupation probabilities: states above the Fermi level become partially occupied, while those below become less populated, effectively broadening the Fermi surface. As a result, new excitation channels become accessible via thermal unblocking of states, particularly in the low-energy region of the E1 response.

To analyze the contribution of PDS across different Ni isotopes and temperatures within the energy region $E = 0$--12~MeV, the $S_{\text{EWS}}$ values are presented in Fig.~\ref{fig3}(a) as a function of temperature for each isotope, and in Fig.~\ref{fig3}(b) as a function of neutron number ($N$) at selected temperatures ($T = 0$--2~MeV). In both panels, the $S_{\text{EWS}}$ values remain nearly constant up to $T = 0.4$~MeV, indicating minimal thermal effects in this low-temperature regime. However, above this threshold, noticeable kinks emerge in the temperature dependence, corresponding to the critical temperature ($T_c$) at which pairing correlations are quenched in open-shell nuclei (see Table~\ref{table1}). A pronounced increase in $S_{\text{EWS}}$ values is observed around $T_c$ for most isotopes, attributable to the onset of thermal unblocking and the collapse of pairing correlations. The case of $^{56}$Ni is particularly noteworthy due to its doubly magic nature and the absence of PDS at zero temperature. As shown in Fig.~\ref{fig3}(a), the value of $S_{\text{EWS}}$ in the low-energy region remains negligible up to $T \approx 0.6$~MeV, followed by a noticeable increase at higher temperatures, driven by the thermal unblocking of previously inaccessible states.

Fig.~\ref{fig3}(b) also illustrates a systematic increase in $S_{\text{EWS}}$ with both neutron number and temperature up to $N = 34$, peaking at $^{62}$Ni. Beyond this point, and after $T=$ 1 MeV, the rate of increase in $S_{\text{EWS}}$ slows between $N = 36$ and $N = 42$, suggesting a saturation of low-energy dipole transitions in heavier isotopes at high temperatures. Notably, $^{62}$Ni exhibits the highest $S_{\text{EWS}}$, reaching approximately 5\% at $T = 2$~MeV, making nuclei in this region the most promising candidates for experimental investigations of HPDR. Although the values of $S_{\text{EWS}}$ generally increase with temperature throughout the Ni isotopic chain, the enhancement is less pronounced in $^{68}$Ni and $^{70}$Ni, as shown in Fig.~\ref{fig3}(b). This reduced sensitivity reflects the influence of proximity to shell closures, which limits the thermal population of higher single-particle states, and consequently restricts the number of available transition channels for dipole excitations at finite temperatures. As a result, thermal unblocking mechanisms play a less significant role in enhancing low-energy dipole strength in these cases compared to other Ni isotopes. Overall, we observe a pronounced enhancement of the pygmy dipole strength in neutron-rich Ni isotopes as temperature increases. In particular, at \( T = 2\,\text{MeV} \), the integrated strength in the low-energy region (\( E = 0\text{--}12\,\text{MeV} \)) increases by up to a factor of 2.5 compared to the zero-temperature case.

The preliminary experimental results by Wieland et al.~\cite{wieland2024extra,wieland2025extra} indicate that the HPDS in the $^{62}$Ni nucleus accounts for approximately 4\% of the TRK energy-weighted sum rule (EWSR) at $T \approx 1.6$~MeV. This is in excellent agreement with our FT-RQRPA calculations, which yield $S_{\text{EWS}} = 4.32\%$ at the same temperature within the energy interval $E = 8$--12~MeV. Additional preliminary experimental findings ~\cite{Wieland2023,Wieland2024}, suggest that the $S_{\text{EWS}}$ values exhaust between 1\% and 8\% of the total GDR-EWSR across Ni isotopes at temperatures up to 2~MeV. These results demonstrate the capability of FT-RQRPA calculations to reliably predict thermally induced dipole strength and serve as robust theoretical input for ongoing and future experimental studies.
To facilitate such comparisons, here we also provide as supplementary material tables reporting TRK-normalized $S_{\text{EWS}}$ values and cumulative $B$(E1) strengths (in e$^2$fm$^2$) for the isotope chain from $^{56}$Ni to $^{70}$Ni. These are tabulated 
from 0 to 12 MeV in 2 MeV increments, with additional bins for 8–12 MeV and for the full 0–12 MeV range, at temperatures ranging from $T = 0$ to 2~MeV.

\begin{figure*}[htp]
\includegraphics[width=\linewidth,clip=true]{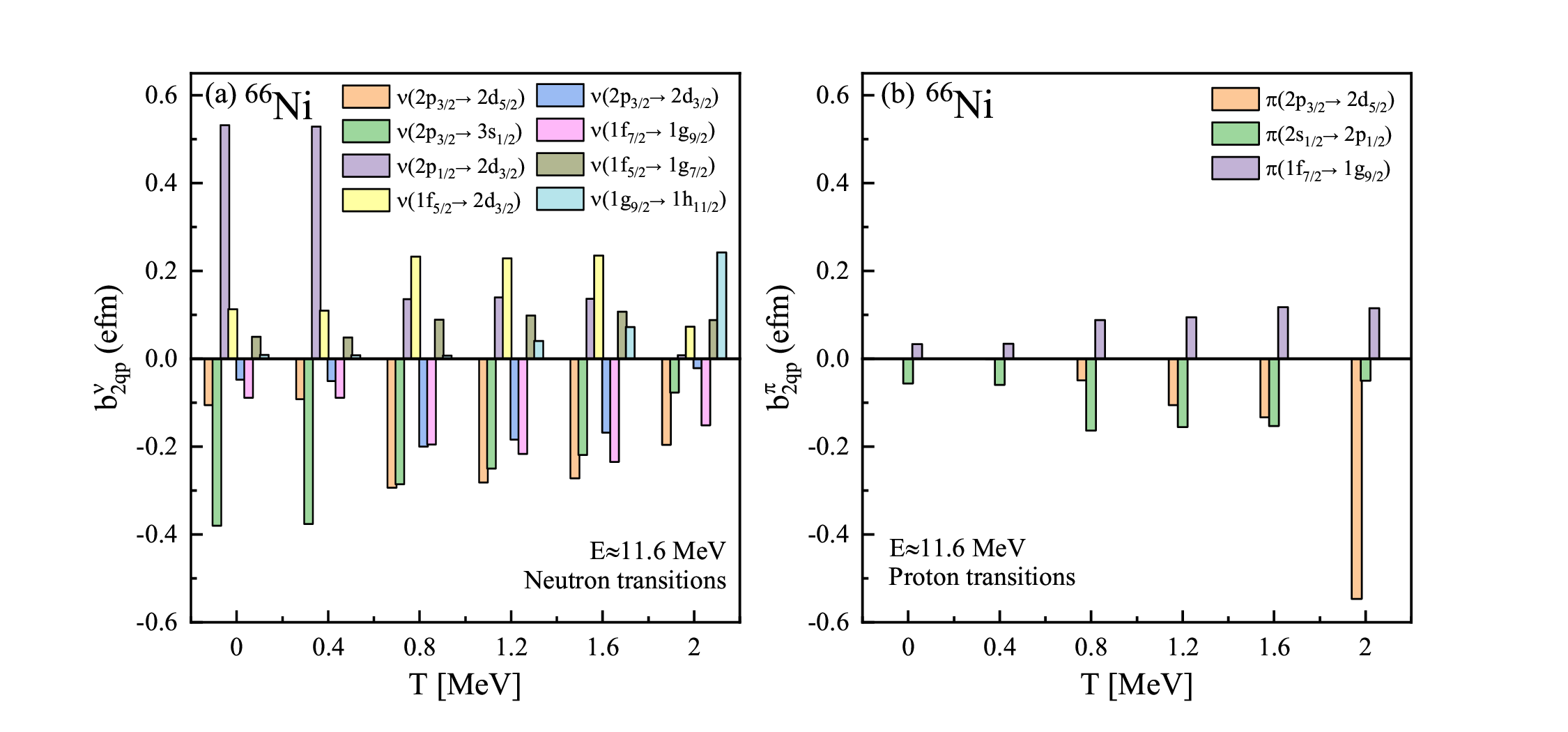}
  \vspace{-0.8cm}
  \caption{ Panels (a) and (b) display the partial contributions $b_{2qp}^{\nu(\pi)}$ of major neutron ($\nu$) and proton ($\pi$) transitions, respectively, to the total isovector E1 strength associated with the PDS state at an excitation energy of approximately 11.6 MeV, as a function of temperature for $^{66}$Ni nucleus. }
  \label{fig4} 
\end{figure*}

To gain a deeper understanding of the underlying physics of the HPDS and its evolution with temperature, it is useful to analyze the contributions of individual two-quasiparticle ($2qp$) configurations to the total transition strength $B$(E1) for excited states at a given energy. Figures ~\ref{fig4}(a) and \ref{fig4}(b) show the temperature dependence of the partial contributions of the neutron ($b_{2qp}^{\nu}$) and proton ($b_{2qp}^{\pi}$) $2qp$ configurations, calculated using Eq.~(\ref{Part.Contri}). These contributions correspond to the total isovector E1 strength associated with the PDS state at $E \approx 11.6$~MeV in the neutron-rich $^{66}$Ni nucleus. 
The reduced transition probability $B$(E1, $E$) for a specific excited state is obtained by summing the contributions of individual configurations, taking into account their relative signs.
The following findings are evident from the figure:
(i)An inspection of the partial E1 strengths at the PDS peak (\( E \approx 11.6\,\text{MeV} \)) reveals that neutron configurations dominate and exhibit larger amplitudes than their proton counterparts at low temperatures, while proton excitations begin to play a more significant role at higher temperatures.
(ii) It is also evident that destructive interference between proton and neutron configurations leads to reduced strength and lowers the collectivity of the PDS.
(iii) The number of $2qp$ transitions increases as the temperature increases from $T = 0$ to 2~MeV, indicating that new configurations, such as $\nu(1g_{9/2} \rightarrow 1h_{11/2})$ and $\pi(2p_{3/2} \rightarrow 2d_{5/2})$, emerge at higher temperatures due to thermal unblocking of transitions.
(iv) Increasing temperature alters the contributions of dominant configurations and can lead to either a decrease or an increase in the total $B$(E1, $E$) value for a given state, as thermally induced changes in occupation probabilities modify the configuration space. For instance, the partial strength of the $\nu(2p_{1/2} \rightarrow 2d_{3/2})$ transition decreases at higher temperatures for E $\approx$ 11.6 MeV. This kind of behavior can be interpreted as a suppression of some configurations at higher temperatures at particular excitation energy, although these configurations may still contribute to other excited states.

\begin{figure}[ht!]
\includegraphics[width=\linewidth,clip=true]{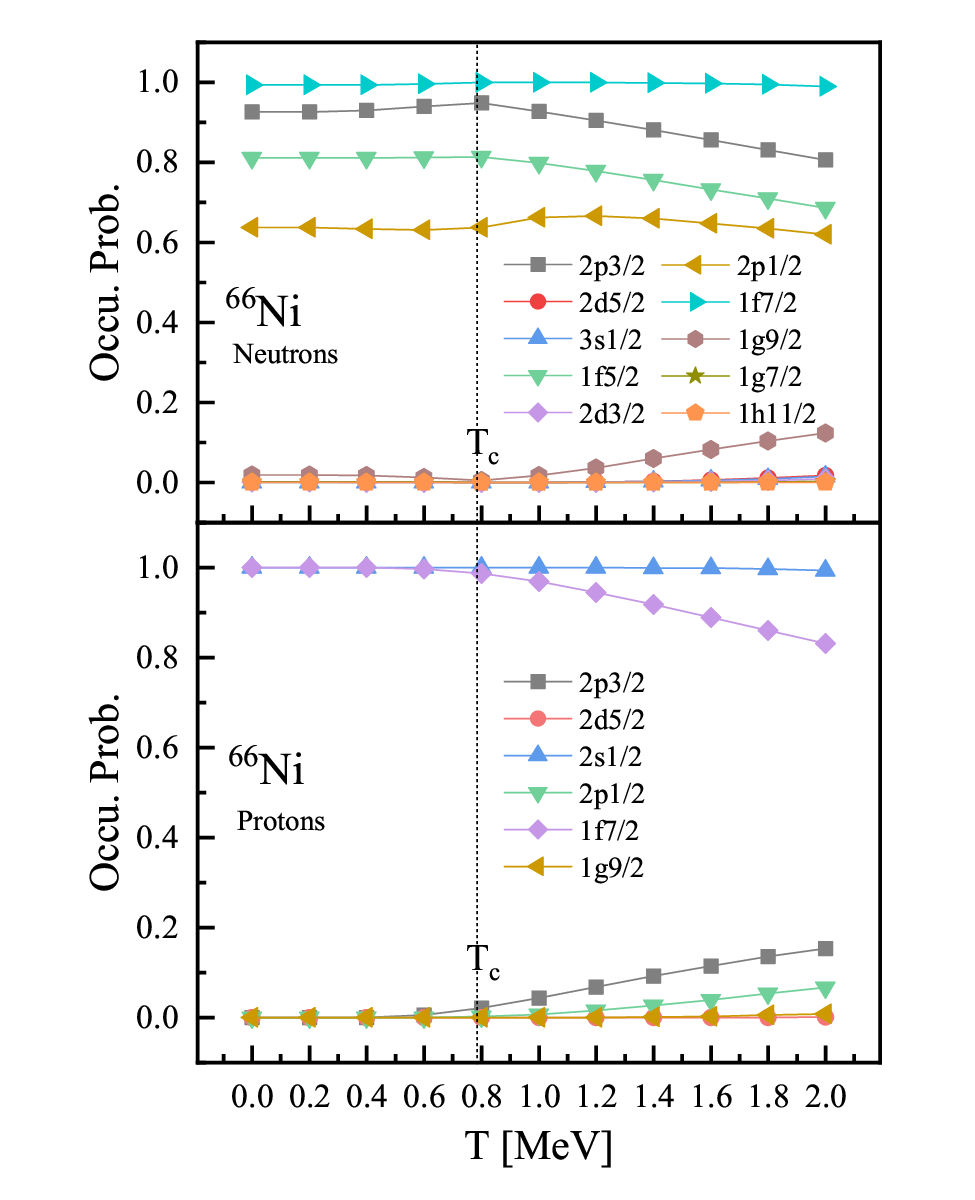}
  \vspace{-0.8cm}
  \caption{Occupation probabilities of neutrons (upper panel) and protons (lower panel) states for $^{66}$Ni as a function of temperature ($T$). }
  \label{fig5} 
\end{figure}

To investigate these effects in more detail, the occupation probabilities of the neutron and proton single-particle states are shown in the upper and lower panels of Fig.~\ref{fig5}, respectively, with emphasis on those contributing to the PDS state at an excitation energy of $E \approx 11.6$~MeV in $^{66}$Ni. As clearly visible in the figure, the neutron and proton occupation probabilities exhibit small changes up to the critical temperature $T_c=0.78$ MeV. Beyond this point, where pairing correlations vanish, the occupation probabilities change significantly. In the upper panel, at low temperatures, the lower-lying neutron orbitals such as $1f_{7/2}$ and $2p_{3/2}$ are nearly fully occupied, while higher orbitals like $1g_{9/2}$, $1g_{7/2}$ and $1h_{11/2}$ remain almost empty. As the temperature increases, a gradual depopulation of the initially occupied orbitals occurs, accompanied by a thermal population of the higher-energy orbitals. In particular, the $1g_{9/2}$ state shows a notable increase in occupancy above $T\approx1$~MeV, indicating enhanced neutron excitations to higher shells. These results are consistent with Fig. \ref{fig4}(a), which shows a significant increase in the partial contribution of $\nu(1g_{9/2}\rightarrow1h_{11/2})$ configuration to the transition strength. This evolution with temperature highlights the significant role of thermal effects in modifying the shell structure and contributes to the emergence of low-energy dipole strength at higher temperatures. 

\begin{figure}[ht!]
\includegraphics[width=\linewidth,clip=true]{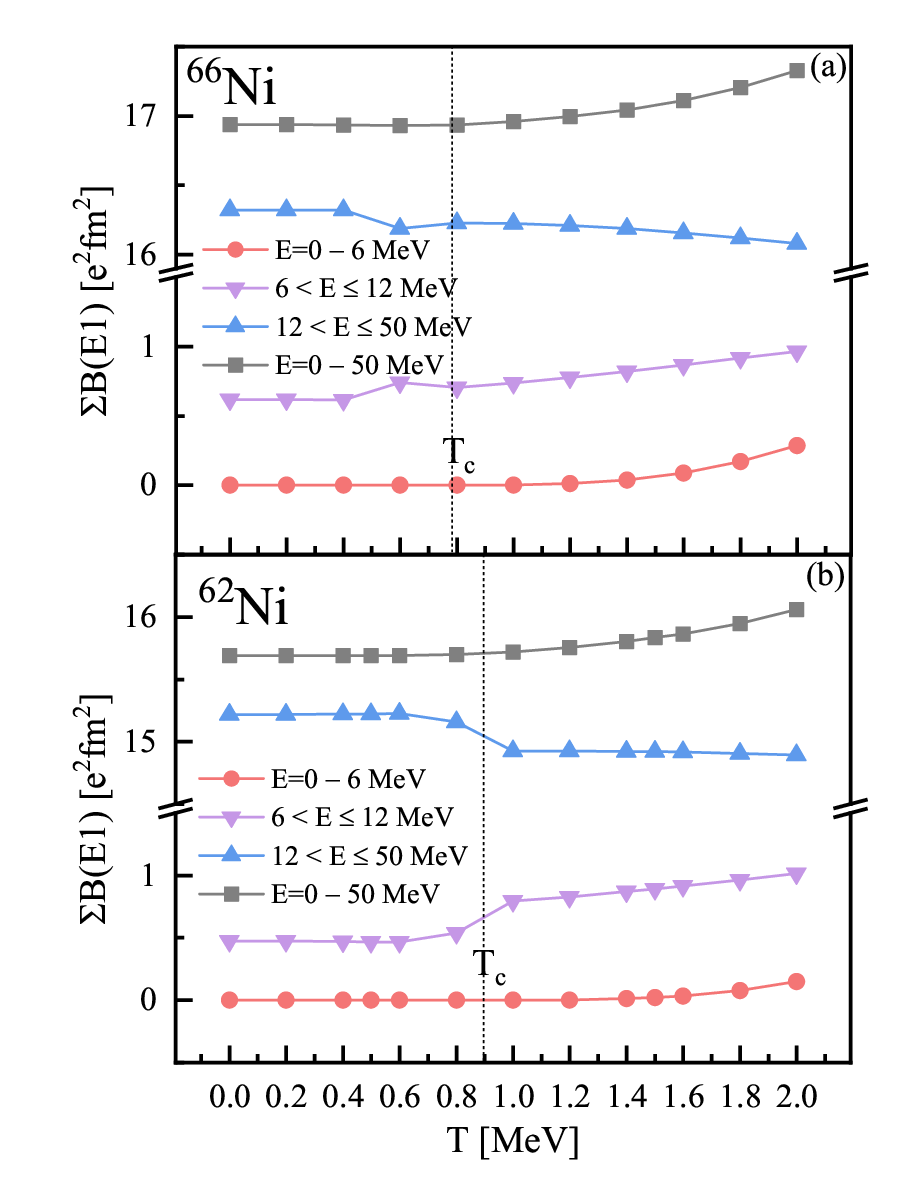}
  \vspace{-0.8cm}
  \caption{Integrated sum of E1 strength at different low-energy region ($E=0-6$ MeV and $6<E\leq12$ MeV MeV) and high-energy region ($12<E\leq50$ MeV and $E=0-50$ MeV) for $^{66}$Ni in upper panel and $^{62}$Ni in lower panel. }
  \label{fig6} 
\end{figure}

For protons, at low temperatures, orbitals such as $1f_{7/2}$ and $2s_{1/2}$ are nearly fully occupied, while higher-lying orbitals like $2p_{3/2}$, $2d_{5/2}$, $2p_{1/2}$ and $1g_{9/2}$ remain largely unoccupied. As the temperature increases, the occupation probabilities of the lower orbitals gradually decrease, while those of the higher orbitals increase, indicating thermal excitations of protons across shell gaps. In particular, the population of the $2p_{3/2}$ orbital rises significantly after $T_c$. The temperature-induced redistribution of occupation plays a crucial role in modifying the proton contribution to the E1 strength at finite temperatures, as evidenced by a sharp increase in the partial contribution of $\pi(2p_{3/2}\rightarrow2d_{5/2})$ transition. This suggests that the interplay between pairing and thermal effects can modify the occupation probabilities of single-particle states. As a result, not only the new excitation channels become accessible due to thermally unblocked states, but the contribution from pre-existing channels may also decrease at higher temperatures.

Figs. \ref{fig6}(a) and \ref{fig6}(b) present the $T$-dependent integrated E1 strength $\Sigma$B(E1) of $^{66}$Ni and $^{62}$Ni, respectively, over distinct energy intervals, including low-energy regions such as $E=0-6$ MeV and $6<E\leq12$ MeV, and high-energy region of $12<E\leq50$ MeV and the full range $E=0-50$ MeV. The total E1 strength over the entire energy range $E=0-50$ MeV shows an increase with increasing temperature for both nuclei. In the energy range $E=0$ to 6 MeV, no significant E1 strength is observed at low temperatures. However, the E1 strength gradually increases as the temperature increases above the critical value, reaching a noticeable enhancement at $T=$ 2 MeV. This $T$-dependent behavior indicates that thermal effects can unblock transitions by modifying occupation probabilities, thereby enabling low-energy dipole excitations that are otherwise suppressed at lower temperatures. Also, beyond the critical temperature $T_c$, the total E1 strength in the low-energy region ($6<E\leq12$ MeV) increases rapidly, whereas the E1 strength in the higher energy region ($12<E\leq50$ MeV) decreases sharply with increasing temperature. The interplay between the high- and low-energy regions indicates a $T$-dependent redistribution of E1 strength, with a portion of the strength from the higher-energy region shifting toward lower energies at higher temperatures. This behavior reflects an enhancement of the pygmy dipole strength as the temperature increases.

\textit{Conclusion.} 
In this work, we employed the recently developed finite-temperature relativistic quasiparticle random phase approximation (FT-RQRPA) to investigate the hot pygmy dipole strength (HPDS) in nickel isotopes with mass numbers A=56-70, across a temperature range of $T=0$ to 2 MeV. The results demonstrate that both temperature and isospin significantly influence low-energy dipole excitations. Low-energy dipole strength is minimal in nuclei with $N\approx Z$ at zero temperature, but becomes significantly enhanced with increasing temperature, indicating the emergence of HPDS. In neutron-rich nickel isotopes, the pygmy dipole strength rises with temperature, reaching values up to 2.5 times larger than those at zero temperature.
A detailed microscopic analysis of the HPDS reveals that the interplay of thermal unblocking and suppression of neutron and proton transitions contribute in a complex manner to the total E1 strength. The redistribution of E1 strength from higher to lower energies with increasing temperature suggests a pronounced enhancement of the pygmy dipole strength in the thermal regime.
The energy-weighted dipole strength ($S_{EWS}$) for low-energy region at finite-temperature, shows good agreement with recently reported preliminary data from the fusion-evaporation experiments \cite{wieland2024extra,wieland2025extra,Wieland2023,Wieland2024}. To support future experiments, we provide supplementary tables containing the calculated $S_{EWS}$ and B(E1) strength values for various low-energy intervals. These results provide an essential theoretical guidance for interpreting ongoing measurements and for shaping future experimental investigations of hot pygmy dipole excitations in thermally excited nuclei.

\textit{Acknowledgements.}
We thank O. Wieland for the insightful discussions on various aspects related to this work. This work is supported by the Croatian Science Foundation under the project Relativistic Nuclear Many-Body Theory in the Multimessenger Observation Era (HRZZ-IP-2022-10-7773). E.Y. acknowledges support from the UK STFC under award no. ST/Y000358/1.

\bibliographystyle{apsrev4-2}
\bibliography{HPDR}
\end{document}